\documentstyle[epsf]{aipproc2}
\begin{document}

\newcommand\beq{\begin{equation}}
\newcommand\eeq{\end{equation}}

\large

\title{Phase Transitions in Dense Matter\\
and Recent Topics of Neutron Stars}

\author{Masatomi Yasuhira}
\address{Department of Physics, Kyoto University,\\
Kitashirakawa-Oiwake-cho,\\
Kyoto 606-8502, Kyoto, JAPAN\\
e-mail:{\it yasuhira@ruby.scphys.kyoto-u.ac.jp}}
\maketitle

\begin{abstract}
Core of neutron star consists of highly dense matter
above normal nuclear density $\rho_0\simeq 0.16{\rm fm}^{-3}$,
where phase transitions is expected
to take place.
We review some phase transitions
and recent topics of neutron stars.
\end{abstract}

\section*{Introduction}
There exist many candidates
of the phase transitions in highly dense nuclear matter:
boson condensation ($K^-$,$\pi$),
quark matter,
hyperonic matter and so on\cite{heiselberg}.
They are often discussed
with respect to neutron star (NS) physics,
for example,
maximum mass,
cooling mechanism,
glitch mechanism,
delayed collapse,
gamma ray burst and
magnetar.

In this paper after the introduction
why we believe the phase transitions
in the core of NS,
We concentrate on three topics:
the delayed collapse due to kaon condensation,
the mixed phase problem for first order phase transitions
and strange stars as magnetar candidates.

\subsection*{Why do we believe the phase transitions?}
The equation of state (EOS) of
normal nuclear matter has been studied
by many authors with various theoretical approaches.
G-matrix and variational calculations
(See ref.\cite{heiselberg} and references therein.)
are based on the microscopic theory,
whose objective is to reproduce the properties of matter
(saturation density, binding energy and nuclear incompressibility)
based on the experimental data(2- or 3-body interaction).
On the other hand,
relativistic mean field theory
is the effective theory
and seems to be very useful method.
The numerical table of EOS is submitted
by Shen et al\cite{shen}(also refer to talk by Prof. Sumiyoshi).

There has been suggested phase transitions
in NS; e.g. nucleon or quark superfluidity,
pion or kaon condensation,
deconfinement transitions.
In this paper we discuss the phase transitions
beyond the normal matter
by the following reasons.
We show two reasons why we need some phase transition:
maximum mass and cooling mechanism.

\subsubsection*{Maximum Mass}
\begin{figure}[ht]
 \vspace{2mm}
 \begin{minipage}{0.49\textwidth}
  \epsfsize=0.99\textwidth
  \epsffile{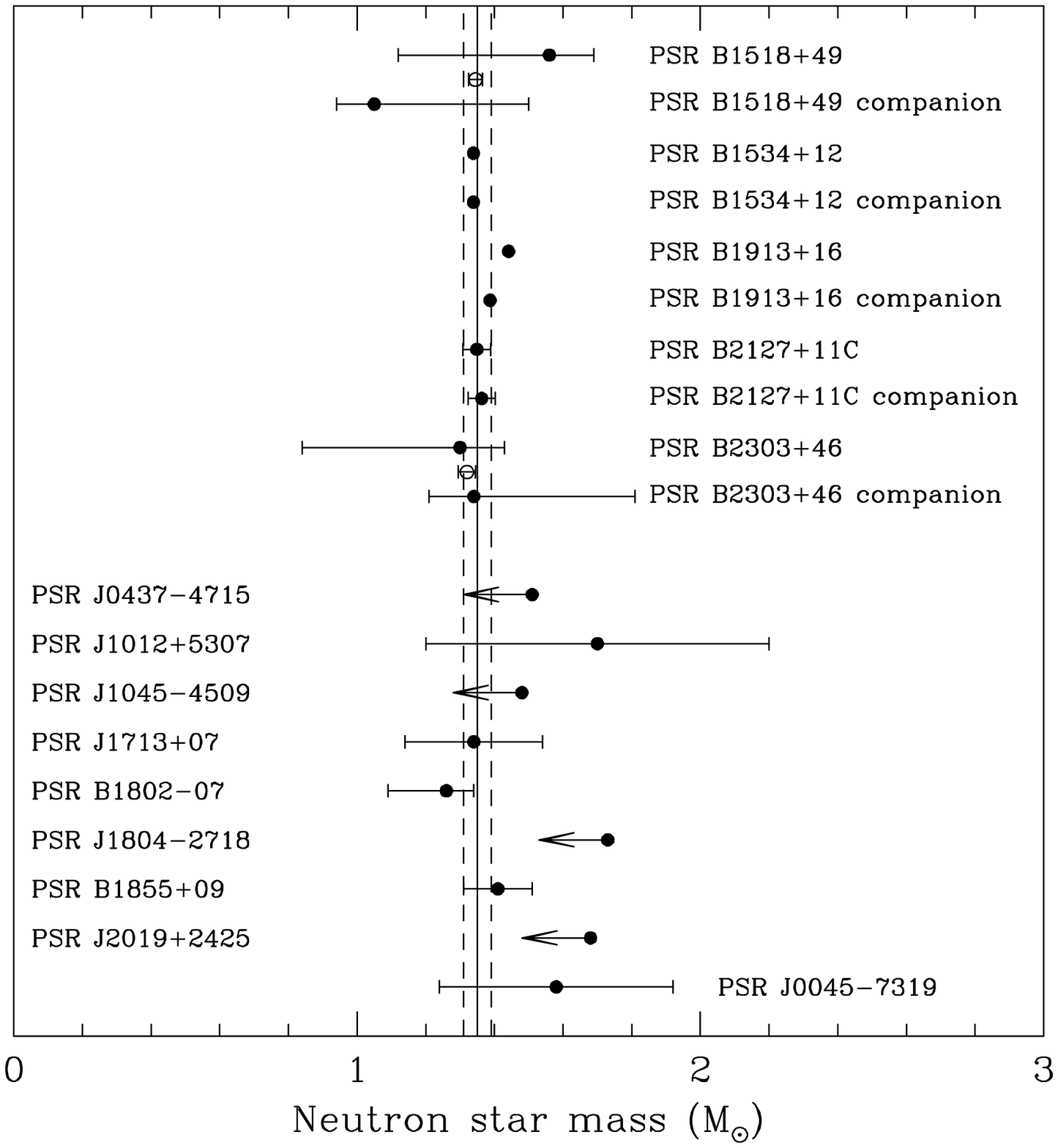}
  \caption{Observed masses of NS from radio pulsar systems
taken from ref.[3].
The vertical lines are drawn at $1.35\pm 0.04M_\odot$.}
  \label{fig:masses}
 \end{minipage}%
 \hfill~%
 \begin{minipage}{0.49\textwidth}
  \epsfsize=0.99\textwidth
  \epsffile{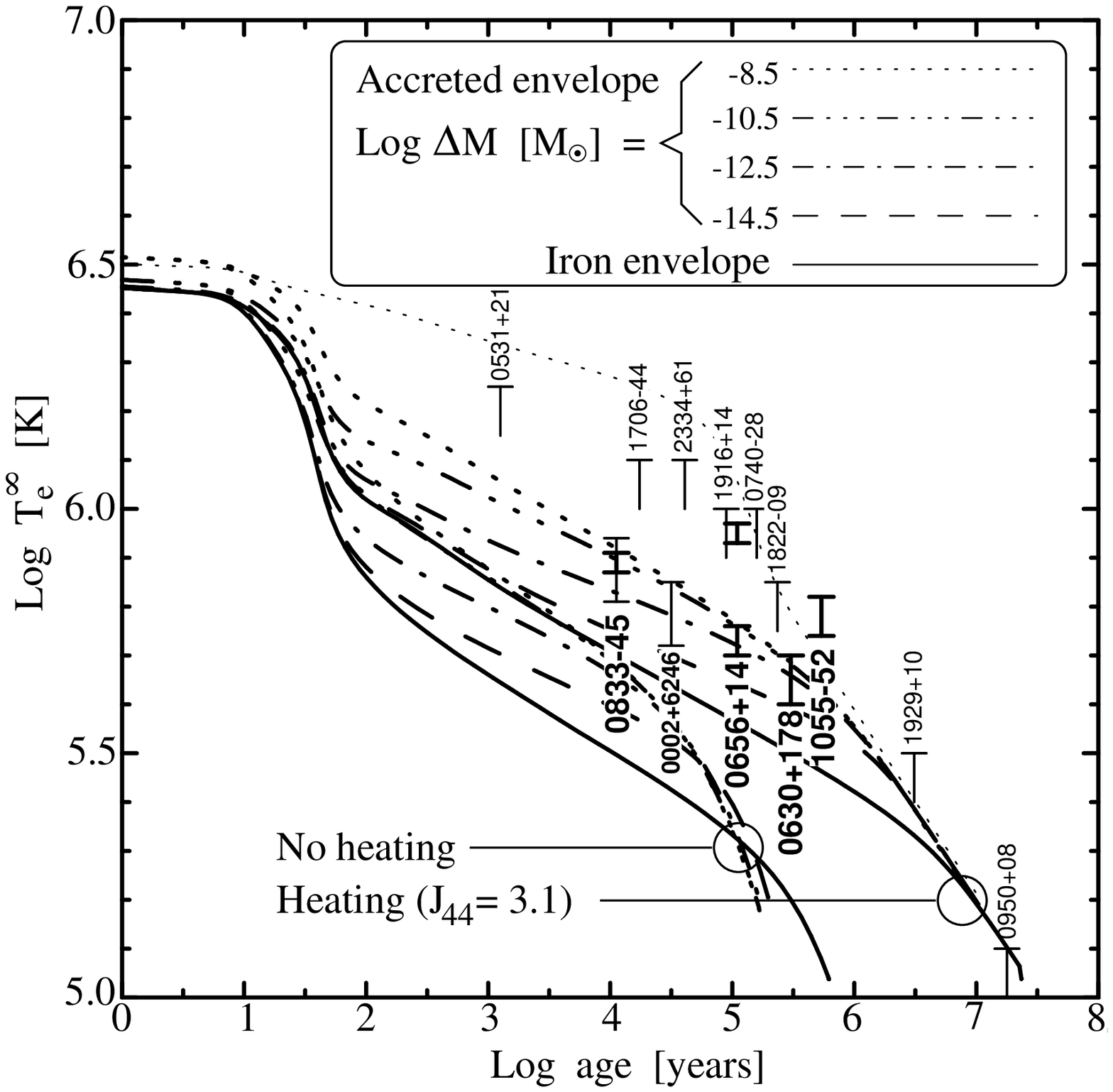}
  \caption{Cooling curves of NS
in the case of standard cooling,
fast cooling due to kaon condensation,
and resulting one with heating.
Taken from ref.[8]}
  \label{fig:cool}
 \end{minipage}%
\end{figure}

Using the EOS of the normal matter,
maximum mass of NS is evaluated about $2.0M_\odot$.
But $M\simeq 1.35M_\odot$ from the observation.
In Fig.\ref{fig:masses}
observed data published by Thorsett, et al. is shown.
There is a large difference
between observation and theoretical calculation.
On the other hand,
if there exists
some phase transition,
for example, kaon condensation,
$M_{max}\simeq 1.4-1.6M_\odot$.
and it seems to be more plausible\cite{thorsson,fujii}.

\subsubsection*{Cooling Mechanism}
In the cooling scenario,
the main contribution in the normal matter
is the modified URCA reactions.
\begin{eqnarray}
 && n + n \to n + p + e^- + \bar{\nu_e}, \nonumber \\
 && n + p + e^- \to n + n + \nu_e. \nonumber
\end{eqnarray}
This scenario based on the normal matter
is called standard cooling scenario,
and it has been suggested that
we need extra cooling mechanisms
to reproduce the observational data points
consistently\cite{Tsuruta,Vosk}.
Each phase transition leads
to the additional rapid cooling mechanisms.
(The cooling curve including these mechanisms
is called the non-standard scenario.)
In the case of kaon condensation,
$K$-induced URCA process exists,
\begin{eqnarray}
 && n + \langle K^- \rangle \to n + e^- + \bar{\nu_e}, \nonumber \\
 && n + e^- \to n + \langle K^- \rangle + \nu_e, \nonumber
\end{eqnarray}
where $\langle K^- \rangle$ means the condensed
kaon field\cite{Tsuruta}.
The cooling curves for non-standard cooling scenario
show the fast cooling
and with more additional process(heating or pairing),
we can explain the observed surface temperature of NS.
Fig.\ref{fig:cool} by D. Page\cite{page} shows
the cooling curves in both cases
and the resulting cooling curve
seems to fit the observational data.

We hereafter address three current issues
about the phase transitions in neutron stars.

\section{Delayed Collapse}\label{sec:dc}
After supernovae explosions,
protoneutron stars (PNS)
are formed with hot,
dense and neutrino-trapped matter.
They usually evolve to cold ($T\simeq0$) NS
through two main eras:
One is deleptonization era
and the other the initial cooling era.
In the deleptonization era,
trapped neutrinos are released in about several seconds
and PNS evolve to hot and neutrino-free NS.
Then through the initial cooling era,
they evolve to cold usual NS in a few tens of seconds.

However some of them may collapse to low-mass black holes
during these eras by softening the EOS
due to the occurrence of hadronic phase transitions\cite{BrownBethe}.
This is called the {\it delayed collapse}.
As a typical example,
neutrinos from SN1987A were observed at Kaomiokande,
but no pulsar yet,
which suggests the possibility of the delayed collapse
in SN1987A.

\begin{figure}[ht]
 \vspace{2mm}
 \begin{minipage}{0.49\textwidth}
  \epsfsize=0.99\textwidth
  \epsffile{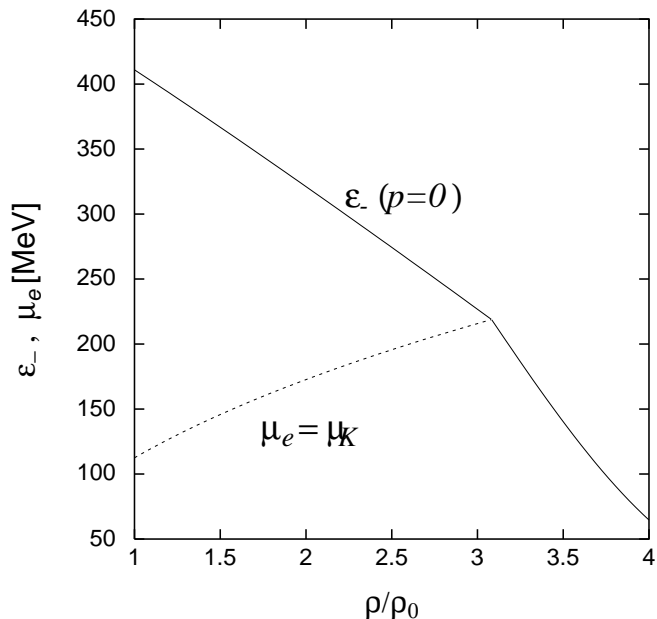}
  \caption{Mechanism of the occurrence of kaon condensation.
Crossing point of chemical potential
and single particle energy for $K^-$.
represents a critical density.}
  \label{fig:disp}
 \end{minipage}%
 \hfill~%
 \begin{minipage}{0.49\textwidth}
decreases due to the attractive $KN$ interaction
in medium,
while the electron chemical potential,
which is equal to the kaon chemical potential
in the beta equilibrium and neutrino-free matter,
increases.
When they become equal to each other,
Bose-Einstein condensation of kaons occurs
at a critical density $\rho_c$.
Kaon condensation
has been studied by many authors
mainly at zero temperature
since first suggested by Kaplan and Nelson\cite{Kaplan}.
We know that the kaon condensation gives rise to
the large softening of EOS.
Kaon condensation is the first order phase transition
and thereby
EOS includes thermodynamically unstable region.
We applied the Maxwell construction to obtain the equilibrium curve
for simplicity
though, restrictly speaking, we need to take the
Gibbs conditions(See Sect.\ref{sec:mp}).
 \end{minipage}%
\end{figure}

We consider here the possibility of the delayed collapse
in the context of kaon condensation.
Kaon condensation,
one of the candidates of the hadronic phase transitions
in high-density nuclear matter,
is a kind of Bose-Einstein condensation.
Fig.\ref{fig:disp} shows the mechanism
of the occurrence of kaon condensation.
As density increases,
kaon's single particle energy $\varepsilon_-$

Recently,
to study the PNS,
there appear a few works
about kaon condensation at finite temperature\cite{Prakash,TandE}
but there was no consistent theory based on chiral symmetry.
Then we have presented a new formulation to treat fluctuations
around the condensate based on chiral 
symmetry\cite{TTMY98l,TTMY}.

With thermodynamic potential in the reference\cite{TTMY98l,TTMY},
we can study the properties of kaon condensed state
at finite temperature
and then discuss some implications 
on the delayed collapse of PNS.
We, hereafter, use the heavy-baryon limit for nucleons\cite{TTMY98l}.
We show the phase diagram, EOS
and then discuss the properties of PNS
where thermal and neutrino-trapped effects
are very important.

\subsection{Phase Diagram and EOS for kaon condensation}
\begin{figure}[ht]
 \vspace{2mm}
 \begin{minipage}{0.49\textwidth}
  \epsfsize=0.99\textwidth
  \epsffile{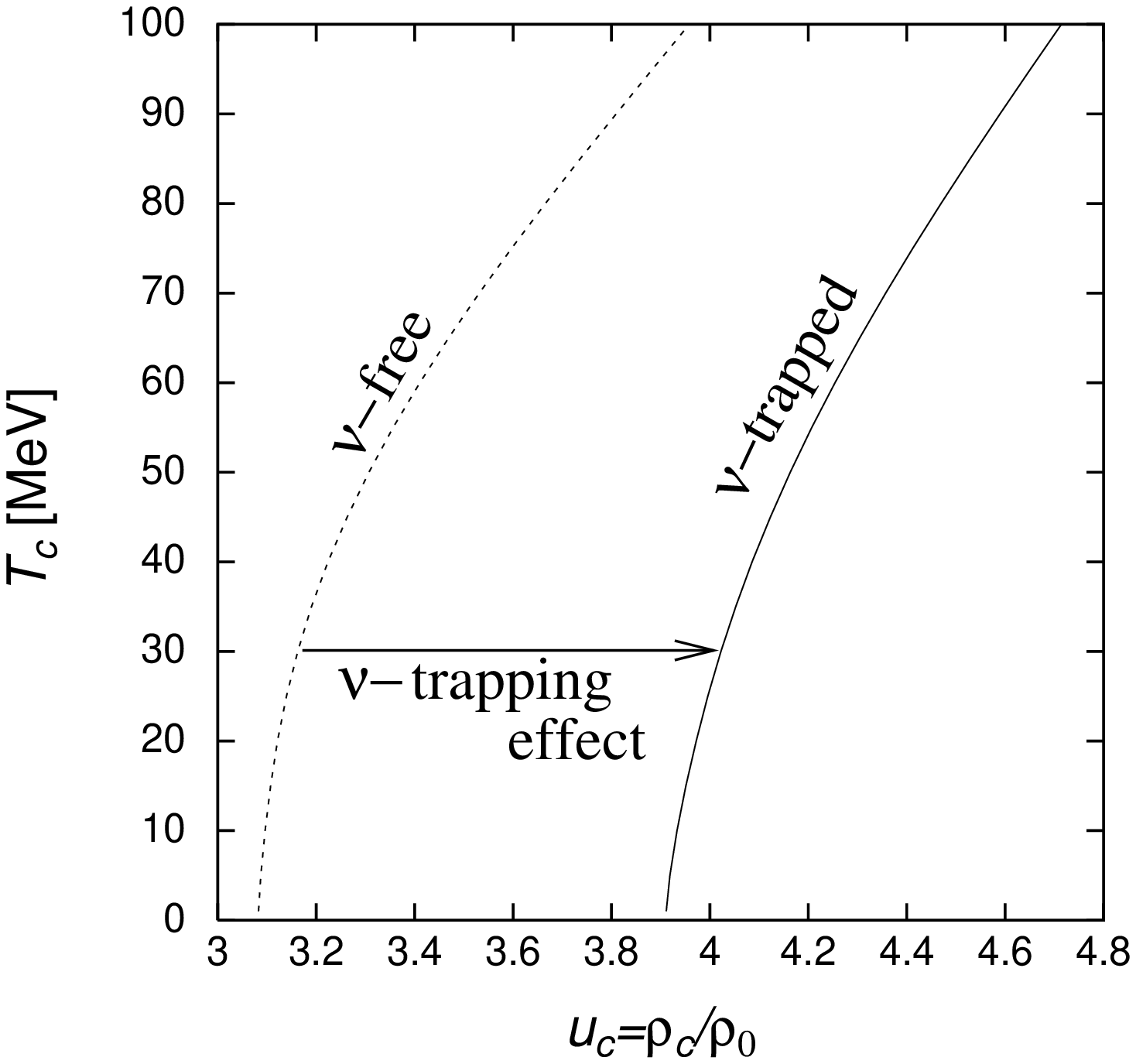}
  \caption{Phase diagram in density and temperature plane.}
  \label{fig:pd}
 \end{minipage}%
 \hfill~%
 \begin{minipage}{0.49\textwidth}
  \epsfsize=0.99\textwidth
  \epsffile{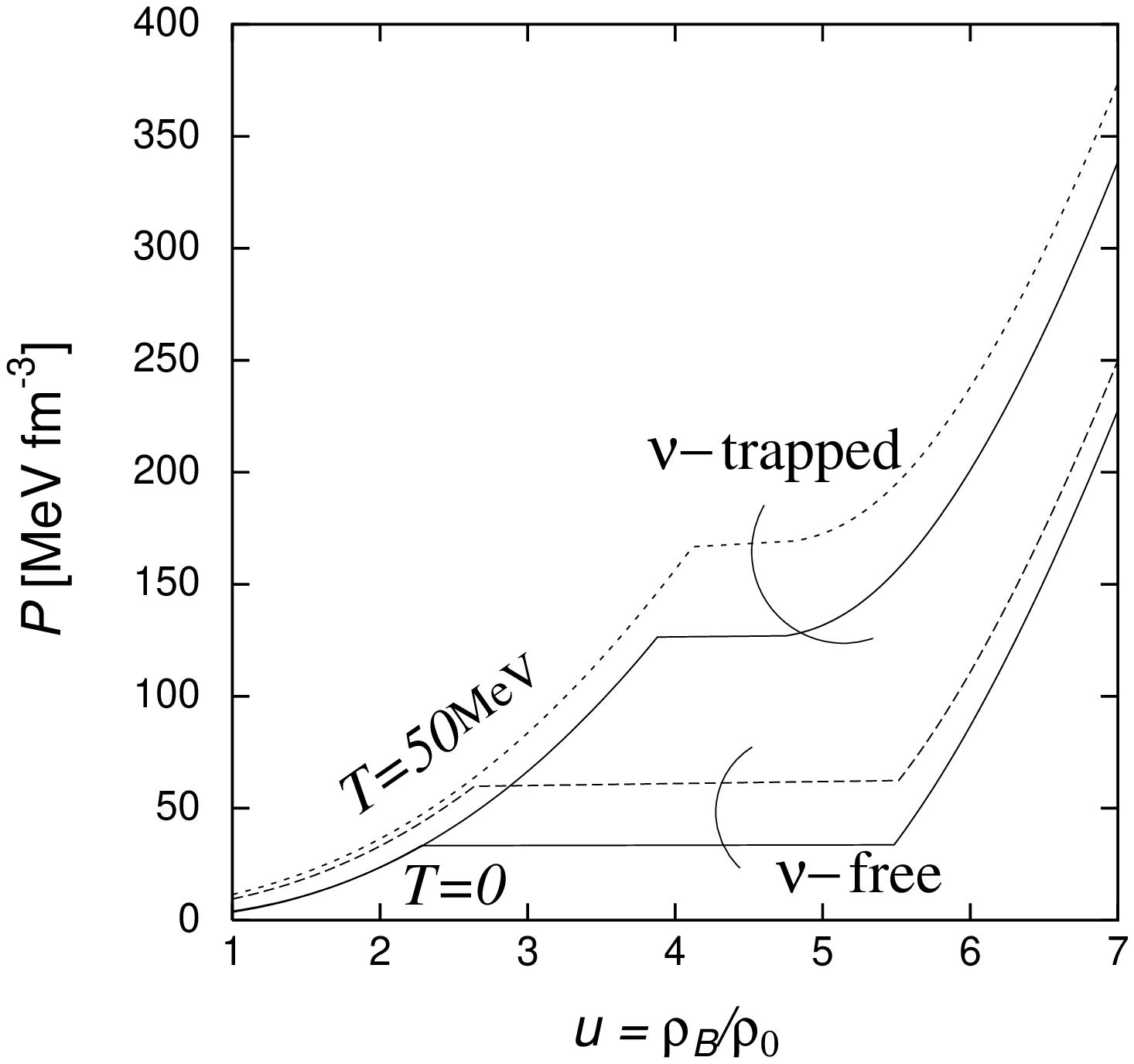}
  \caption{The EOS for kaon condensed matter.}
  \label{fig:eos}
 \end{minipage}%
\end{figure}

First we show the phase diagram 
in Fig.\ref{fig:pd}.
In the neutrino-trapped case
we set $Y_{le}=Y_e+Y_{\nu_e}=0.4$ where
$Y_e$($Y_{\nu_e}$) is the electron(electron-neutrino)
number per baryon,
while $Y_{\nu_e}=0$ in the neutrino-free case.
Both of the thermal and neutrino-trapped effects
largely
suppress the occurrence of kaon condensation.
The reason for the latter case
may be understood from
the threshold condition,
$\varepsilon_-(\rho_c)$
$=$
$\mu_K$.
Chemical equilibrium holds the relation
$\mu_K$
$=$
$\mu_e$
$-$
$\mu_{\nu_e}$:
$\mu_{\nu_e}>0$ in the neutrino-trapped case
while $\mu_{\nu_e}=0$ in the neutrino-free case,
which means kaon chemical potential should
be suppressed in the neutrino-trapped case.
Then
the occurrence of kaon condensation is suppressed
through the suppression of number of kaons.

Next Fig.\ref{fig:eos} represents
the EOS in density-pressure plane.
Both of thermal and neutrino-trapping effects
stiffen the EOS in the condensed phase.
They seem to be more pronounced in the condensed state,
especially around the critical density(see Fig.\ref{fig:eos}),
mainly through the rise of critical density.

In the realistic situation in PNS
the isentropic condition is more relevant\cite{Prakash}.
We reconstruct the isentropic EOS
by evaluating the entropy as a function of temperature.

\subsection{Properties of PNS}
\begin{figure}[htb]
 \vspace{2mm}
 \begin{minipage}{0.48\textwidth}
  \epsfsize=0.99\textwidth
  \epsffile{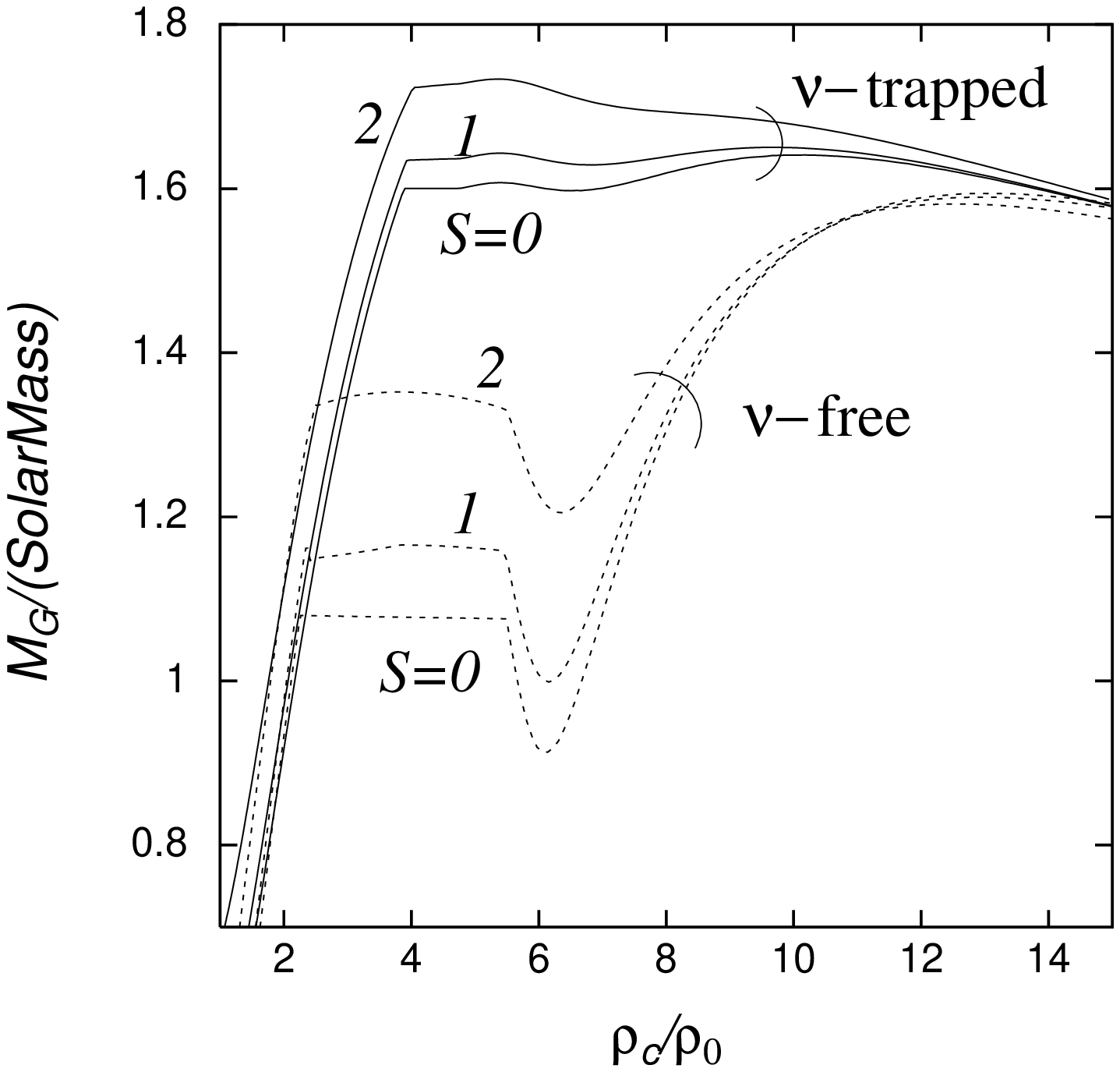}
  \caption{Central density versus gravitational mass.}
  \label{fig:mu}
 \end{minipage}%
 \hfill~%
 \begin{minipage}{0.49\textwidth}
  \epsfsize=0.99\textwidth
  \epsffile{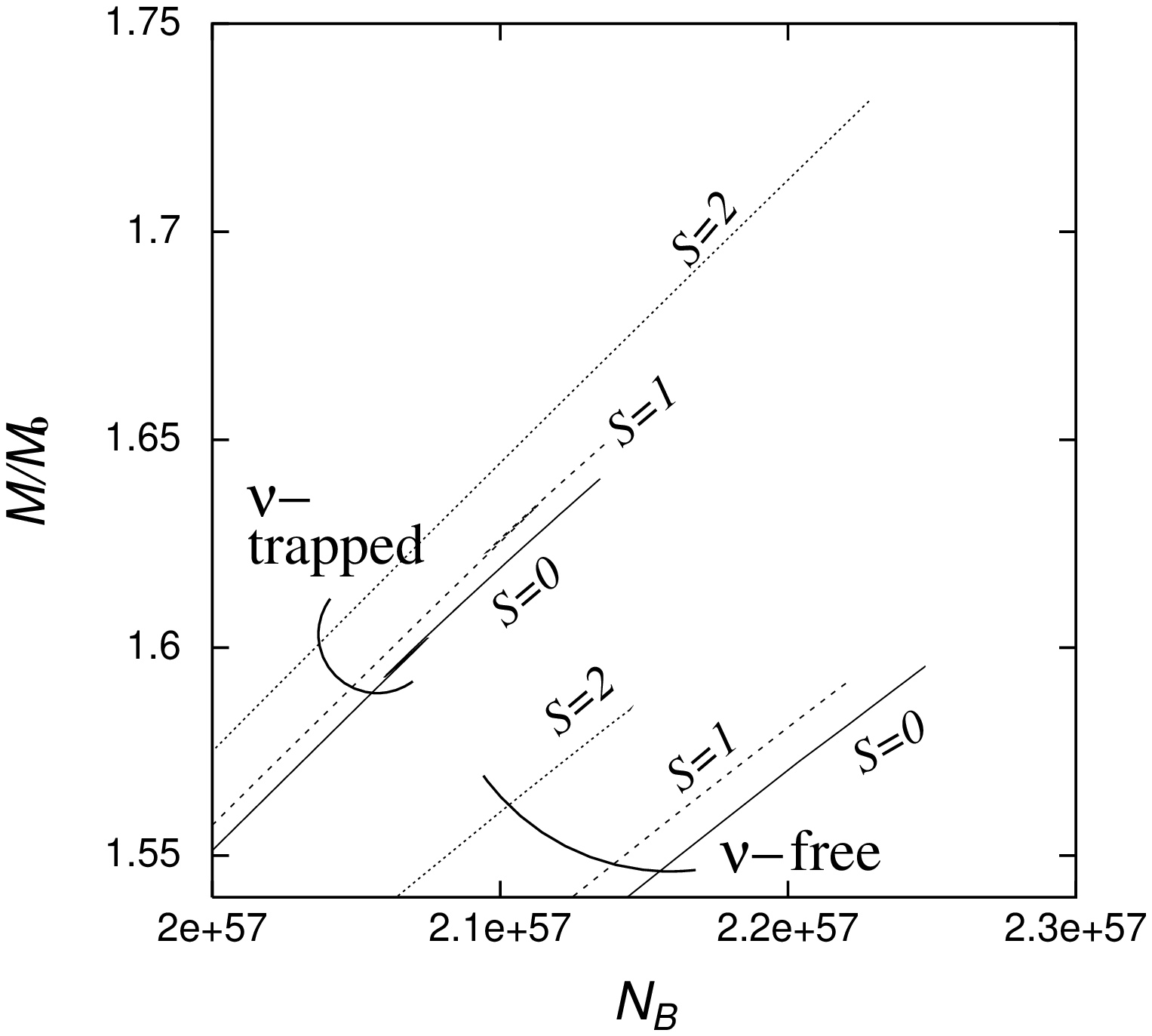}
  \caption{Total baryon number $N_B$ and gravitational mass
for stable PNS.}
  \label{fig:NBM}
 \end{minipage}%
\end{figure}

Solving the TOV equation with the EOS,
we can study the properties of PNS.
In Fig.\ref{fig:mu} we show
the gravitational mass versus central density
for the neutrino-trapped and -free cases
with entropy per baryon $S=0,1$ or $2$.
Both of the thermal and neutrino-trapping effects
make the gravitational mass larger
for the almost all of the central density.
As the exception,
at high central density
in neutrino-free case
masses of hot NS are smaller than ones of cold NS.
This is out of our intuition
and the reason is as follows.
Usually thermal effect enlarges
the pressure and leads to
the larger mass.
At the same time,
however, thermal effect enlarges the energy
which contribute to gravitation.
In the competition of the increase of pressure
and gravitation,
sometimes, mass of hot NS becomes smaller
than one of cold NS.

In the neutrino-free case,
once kaon condensation occurs in the core of a star,
gravitational mass is little changed in the neighborhood
of the equal pressure region in the isothermal EOS,
then
gravitationally unstable region (negative gradient part) appears.
Therefore
the neutron-star branch is separated into two stable branches:
one is for stars with kaon condensate in their cores
(right hand side from gravitationally unstable region)
and the other consisting of only normal matter
(left hand side from gravitationally unstable region).
Thermal effect to the gravitational mass
seems to be very large around the critical density
because the EOS changes largely there.
However the maximum mass,
stars with which include kaon condensate in the core,
is hardly changed 
and even decrease by the thermal effect
as already discussed.
As the temperature raises,
gravitational mass grows largely in the normal branch
but not in the kaon condensed branch.

In the neutrino-trapping case,
thermal effect is large around the critical density
and small for the heavy stars as well,
but the situation is quantitatively different.
For the  $S=0$ or $1$ case
we can see that the neutron-star branch is also separated by the
gravitationally unstable region
and the star with maximum mass
exists in the kaon condensed branch.
(Their central density $\rho_c \simeq 10\rho_0$)
On the other hand,
in the $S=2$ case
almost all of the stars 
with kaon condensate are gravitationally unstable,
and the maximum-mass star whose central density
$\rho_c = 5.4\rho_0$,
still resides in the normal branch.
For this reason
the central density of the maximum-mass star 
is very different from those for $S=0$ or $1$.

To discuss the possibility of delayed collapse of PNS,
the total baryon number $N_B$ should be fixed
as a conserved quantity during the evolution\cite{takatsuka},
under the assumption of no accretion.
In Fig.\ref{fig:NBM}
we show the gravitational mass versus total baryon number for
gravitationally stable PNS omitting unstable stars.
Each terminal point represents
maximum mass and maximum total baryon number.
If an initial mass exceeds the terminal point
in each configuration,
the star should collapse into a black hole
(not a delayed collapse but
a usual formation of a black hole).
We have shown the neutrino-trapped
and -free cases; the former case might be relevant for
the deleptonization era,
while the latter for the initial cooling era.
It is interesting to see the
difference between
the neutrino-free and -trapped cases:
the curve
is shortened as the entropy increases in the former case,
while elongated in the latter case,
where the remarkable increase in $S=2$ and neutrino-trapped case
results from that maximum mass exists
in normal branch,
different from other configurations
with maximum mass in kaon condensed branch.
These features are important for
the following argument about the delayed collapse and
maximum mass of the cold NS.
The delayed collapse is
possible if the initially stable star on a curve
finds
no corresponding point on other curves
during the evolution through deleptonization 
or cooling with the baryon number fixed.
Consider a typical evolution for example:
A PNS is born as a neutrino-trapping and hot ($S=2$) star
after supernova explosion
and evolves
to neutrino-free and hot ($S=2$) stage
through deleptonization.
Then through the cooling,
the star evolves to be neutrino-free and cold ($S=0$).
We can clearly see the PNS with 
large enough mass
can exist as a stable star at the beginning but
cannot find any point on the neutrino-free and $S=2$ curve.
Therefore they must collapse to the low-mass black hole
by deleptonization.
It is to be noted that because the neutrino-trapped and $S=2$ star
never includes kaon condensate,
its collapse is largely due to the appearance of 
kaon condensate in the core.
Thus
we may conclude that
kaon condensation is very plausible to cause the delayed collapse
in the deleptonization era.

On the other hand,
in the initial cooling era after deleptonization
delayed collapse does not take place
because
the NS on neutrino-free and $S=2$ branch
evolve to neutrino-free and cold branch
and all of the stars seem to be able to find
corresponding stable points in the each stage.

\subsection{Summary for delayed collapse}
We have shown that
the delayed collapse is possible 
in the deleptonization era
due to the appearance of kaon condensation
and the maximum mass of cold NS
should be determined
in neutrino-free and hot stage\cite{MYTTdelayed}.

On the other hand,
Pons et al. also studied
kaon condensation in PNS matter\cite{Ponsprep}
They concluded that
the thermal effect is a key object to the delayed collapse
and it is different from ours.
We cannot give a clear reason for
the discripancy at present but
it may originate from the difference of formulations:
our discussion is based on the chiral Lagrangian
and theirs, meson exchange model.
Otherwise the difference may disappear
if they inpose the condition
that the total baryon number is fixed.

In order to study the mechanism of delayed collapse
and mass region which should collapse in more detail,
we had better study the dynamical evolution
beyond the static configurations.
There neutrino opacity is important
to determine the duration of the deleptonization\cite{Pons},
and in the kaon condensed phase
neutrino opacities may become larger
than in normal phase\cite{Reddy,MutoTYI}.
As another remaining issue,
we will refine the EOS
with the Gibbs conditions instead of the Maxwell construction.
 
\section{Mixed Phase}\label{sec:mp}
For first order phase transition,
for kaon condensation, quark matter, etc,
there appears the thermodynamically unstable region
in EOS.
The Maxwell construction,
which may be familiar
for the liquid-gas phase transition for water,
had been used as a standard method
to get the proper EOS.
Recently Glendenning pointed out that
the Maxwell construction is not correct
and we should impose the Gibbs conditions\cite{gle}.

\subsection{Maxwell Construction}
\begin{figure}[ht]
 \vspace{2mm}
 \begin{minipage}{0.32\textwidth}
  \epsfsize=0.99\textwidth
  \epsffile{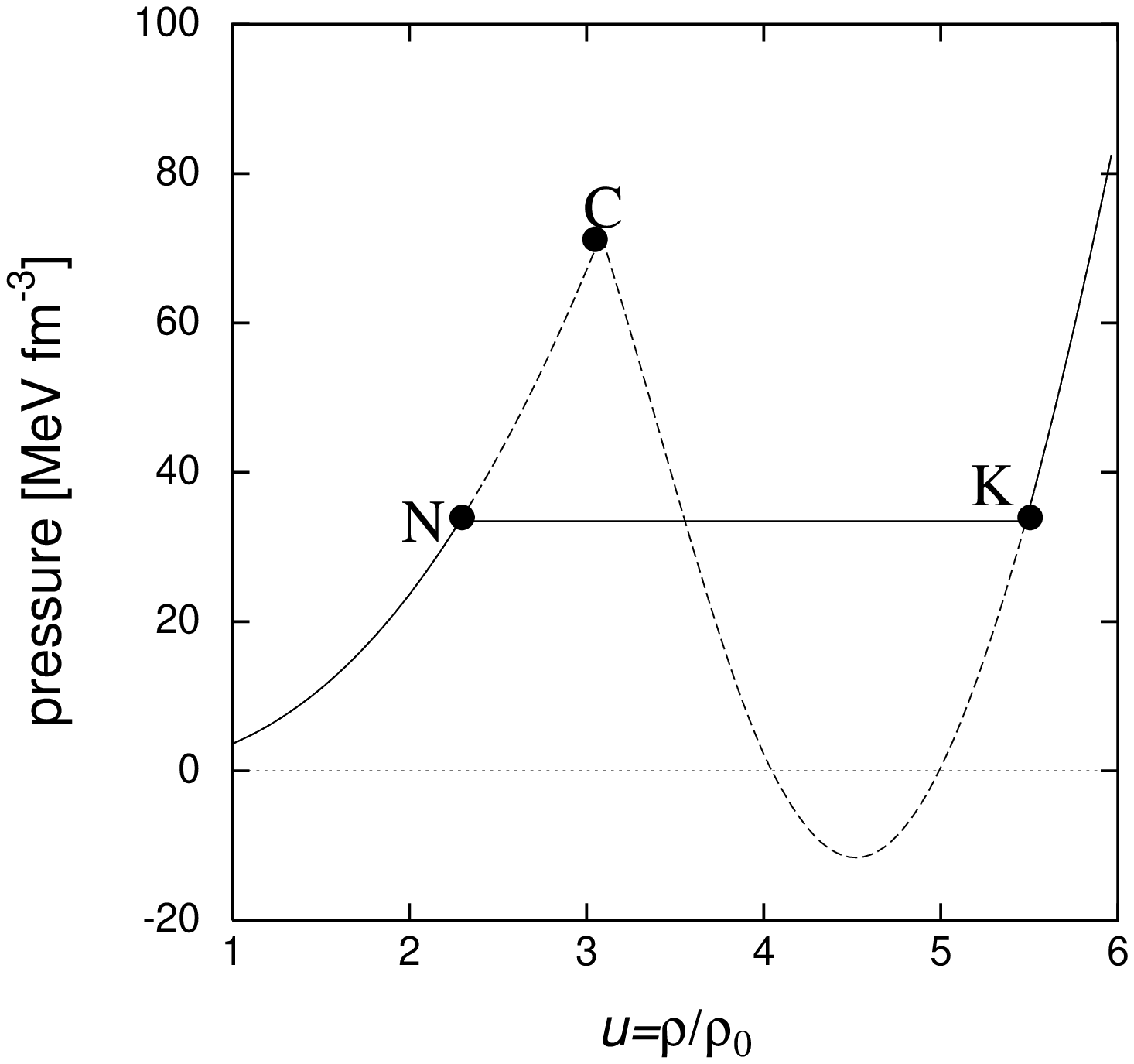}
  \caption{Original EOS and one in the Maxwell construction.
}
  \label{fig:mw}
 \end{minipage}%
 \hspace{0.2cm}
 \begin{minipage}{0.32\textwidth}
  \epsfsize=0.99\textwidth
  \epsffile{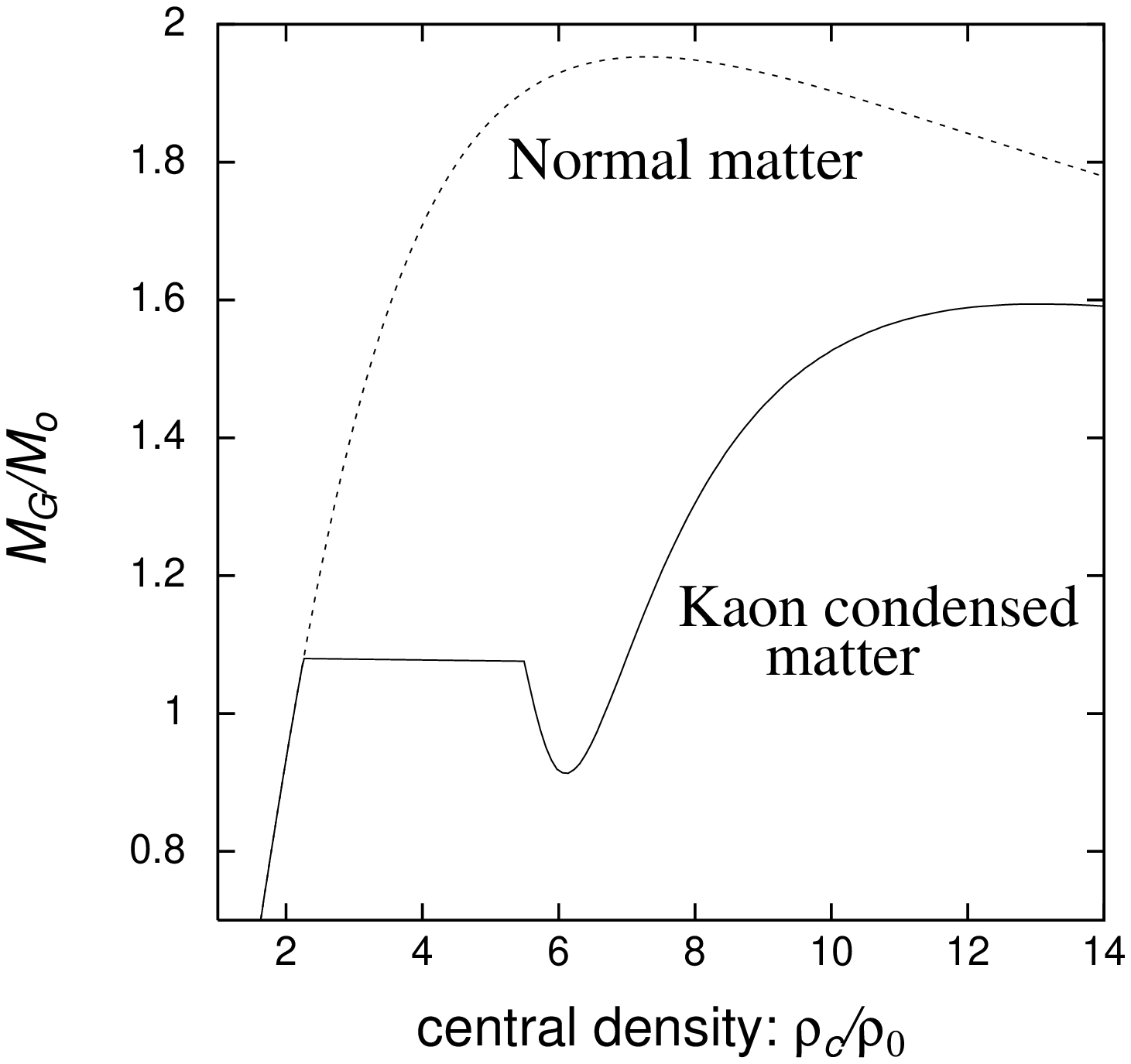}
  \caption{Mass-central density curve.}
  \label{fig:nkmu}
 \end{minipage}%
 \hfill~%
 \begin{minipage}{0.32\textwidth}
  \epsfsize=0.99\textwidth
  \epsffile{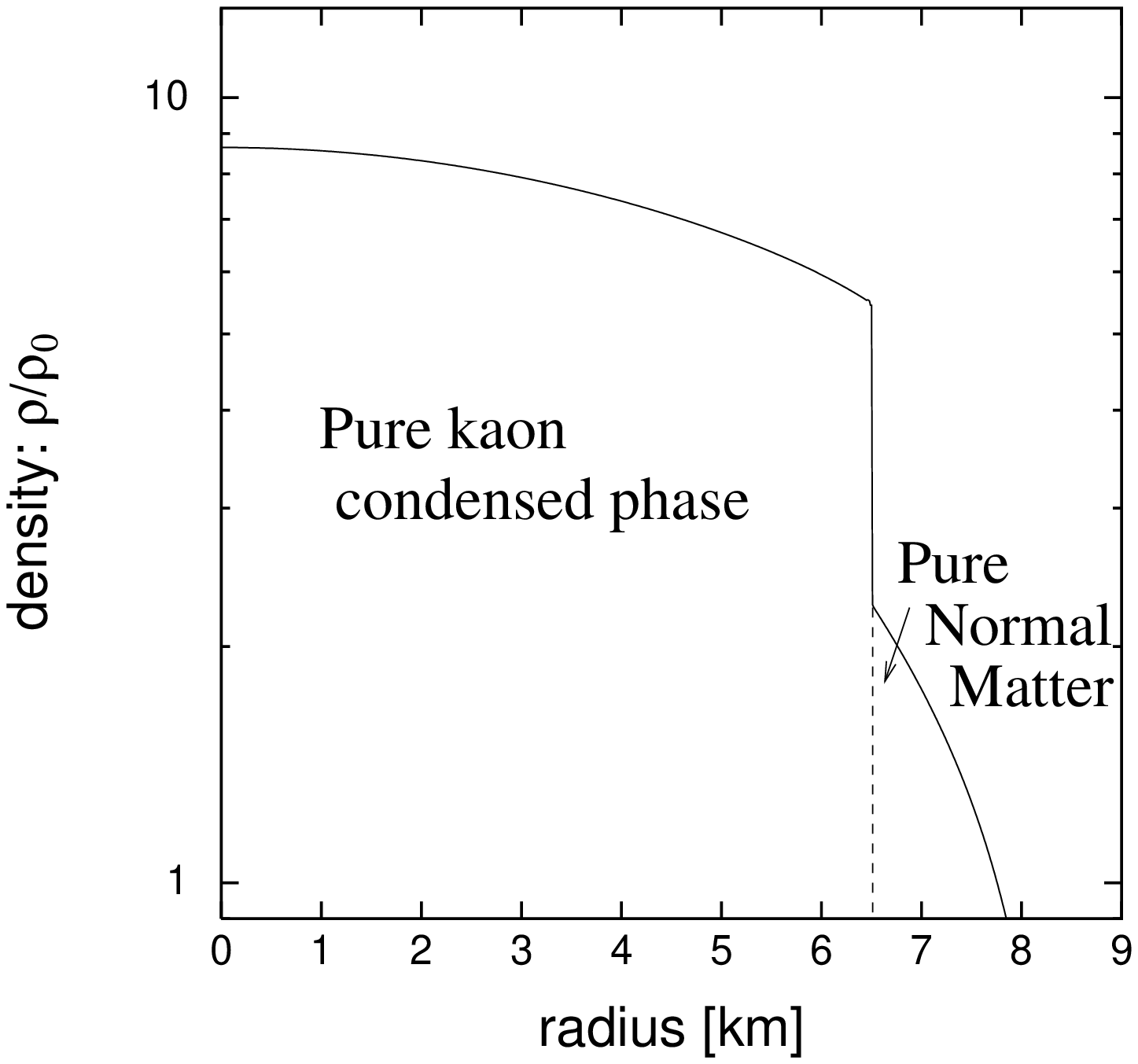}
  \caption{Interior structure of NS.}
  \label{fig:kst}
 \end{minipage}%
\end{figure}

The Maxwell construction imposes
the conditions: $\mu_n^N=\mu_n^K$ 
and $P^N=P^K$,
and can be written as equal-area rule,
\beq
 \int^K_N V dP = 0.
\eeq
In Fig.\ref{fig:mw}
we show the original EOS
with thermodynamically unstable region
and improved one by the Maxwell construction.
As a result of the Maxwell construction,
there appears equal pressure region,
which is called the mixed phase
of kaon condensed matter and normal matter,
only the volume ratio of two phases changes
and properties of each phase
(density, chemical potentials and so on.)
never change.
And the density gap
leads to peculiar structure of NS.
In Fig.\ref{fig:nkmu}
mass-central density curve is shown.
Once kaon condensation occurs at $\rho\simeq 2.3\rho_0$,
density gap appears
and gravitationally unstable region
(region with negative slope) follows.
Then picking up a NS,
the core consists of
two pure phases without the mixed phase (See Fig.\ref{fig:kst}),
which is known as a peculiar feature
due to the Maxwell construction.

\subsection{Gibbs Conditions}
In fact,
however,
for the first order phase transitions in nuclear matter,
the Maxwell construction is not valid.
Because there exist two chemical potentials,
baryon and charge chemical potentials,
we should use the Gibbs conditions\cite{gle}.

The Maxwell construction
is the same to the Gibbs conditions
in case with
only one chemical potential.
On the other hand,
there exist two chemical potentials
in nuclear matter:
baryon and charge chemical potentials
which correspond to neutron and electron chemical potentials
respectively.
Then we need to use the Gibbs conditions,
\beq
 \mu_n^N = \mu_n^{K},
\quad \mu_e^N = \mu_e^{K},
\quad P^N = P^{K},
\eeq
between the phases of normal matter $N$ and
kaon condensed matter $K$.

\begin{figure}[ht]
 \vspace{2mm}
 \begin{minipage}{0.54\textwidth}
  \epsfsize=0.99\textwidth
  \epsffile{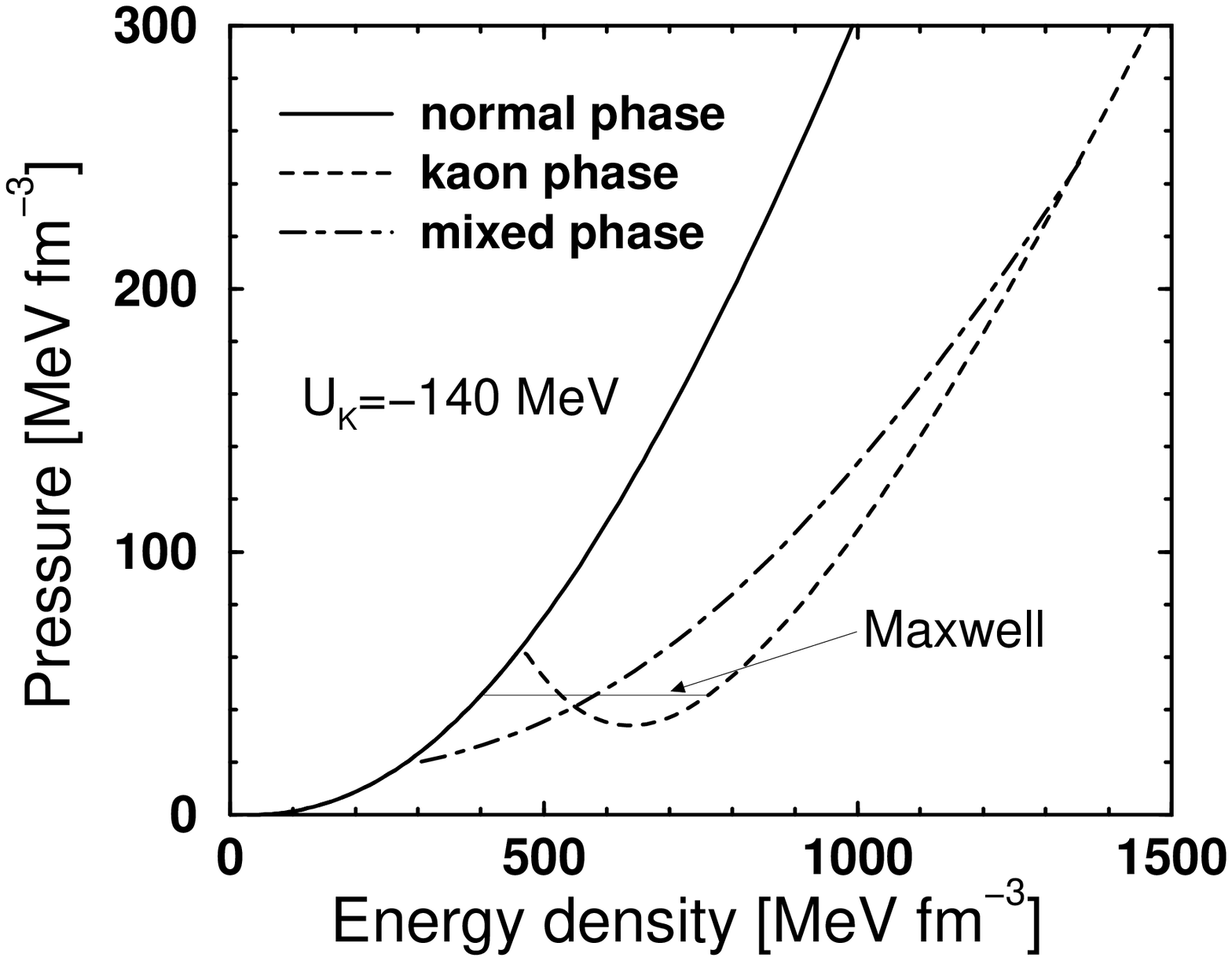}
  \caption{Comparison of EOS taken from ref.[22].
Parameter ${\rm U}$ represents the
kaon's potential contribution.
}
  \label{fig:gleeos}
 \end{minipage}%
 \hfill~%
 \begin{minipage}{0.44\textwidth}
  \epsfsize=0.99\textwidth
  \epsffile{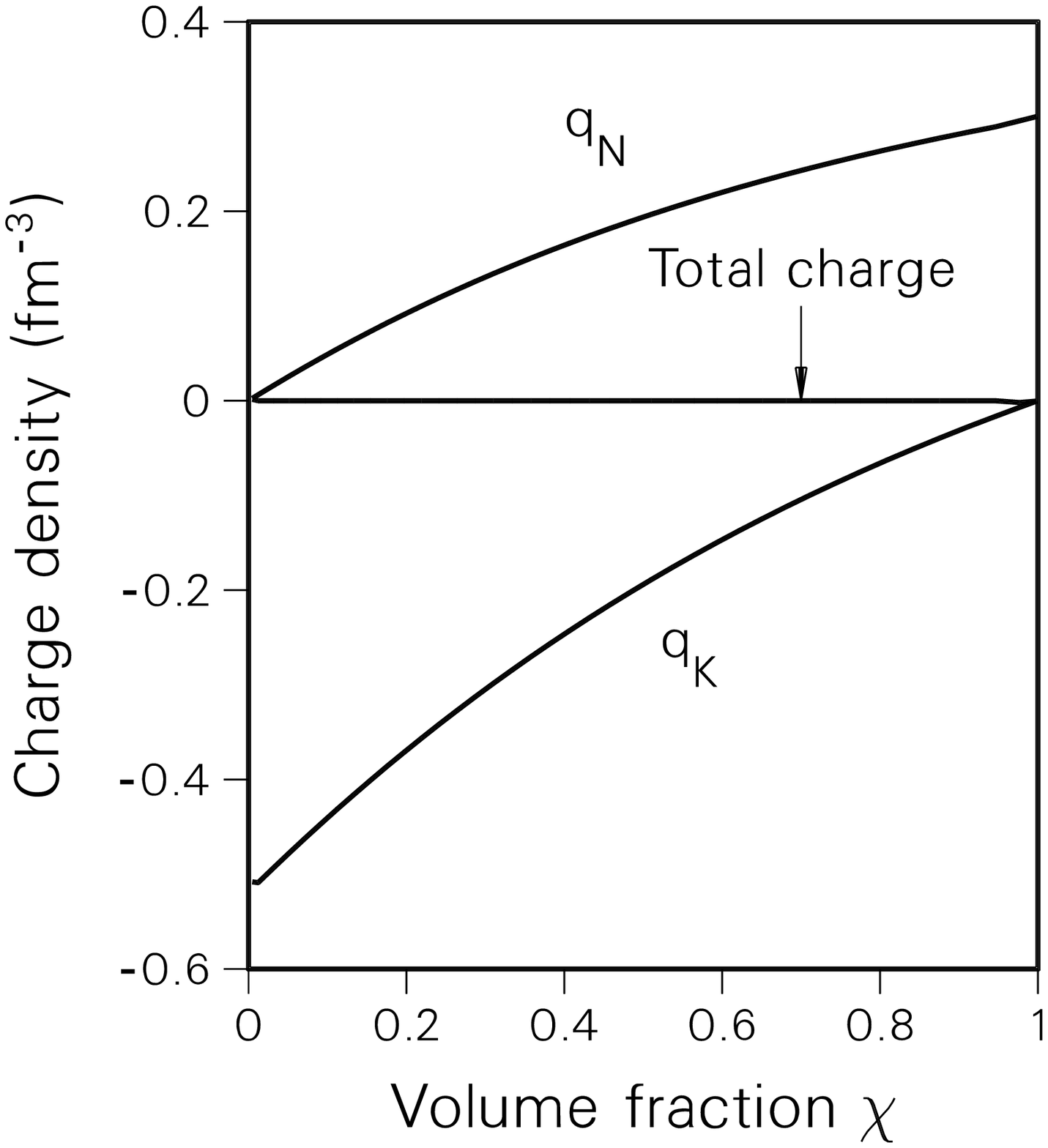}
  \caption{Charge density in mixed phase.
Taken from ref.[22].}
  \label{fig:charge}
 \end{minipage}%
\end{figure}

Glendenning and Schaffner-Bielich discussed
kaon condensation imposing the Gibbs conditions\cite{gleK}
instead of the Maxwell construction.
In Fig.\ref{fig:gleeos},
their result is shown.
In the case of the Gibbs conditions,
compared to the case of the Maxwell construction,
the mixed phase exist in the broad range of density
and the equal pressure region disappears.
The densities for each phase can change
at each point of the mixed phase.
Charge density is nonzero in each phase
and charge neutrality is not achieved locally
(Of course, globally achieved.).
This may be unfamiliar situation
but simply understood as follows.
In the mixed phase
if there exists only baryon chemical potential,
which is in the case of the Maxwell construction,
baryon densities are not equal in two phases.
On the other hand,
in the case there exist two chemical potentials,
as in the case of the Gibbs conditions,
both of baryon and charge densities
are not equal.

\begin{figure}[htb]
 \vspace{2mm}
 \begin{minipage}{0.49\textwidth}
  \epsfsize=0.99\textwidth
  \epsffile{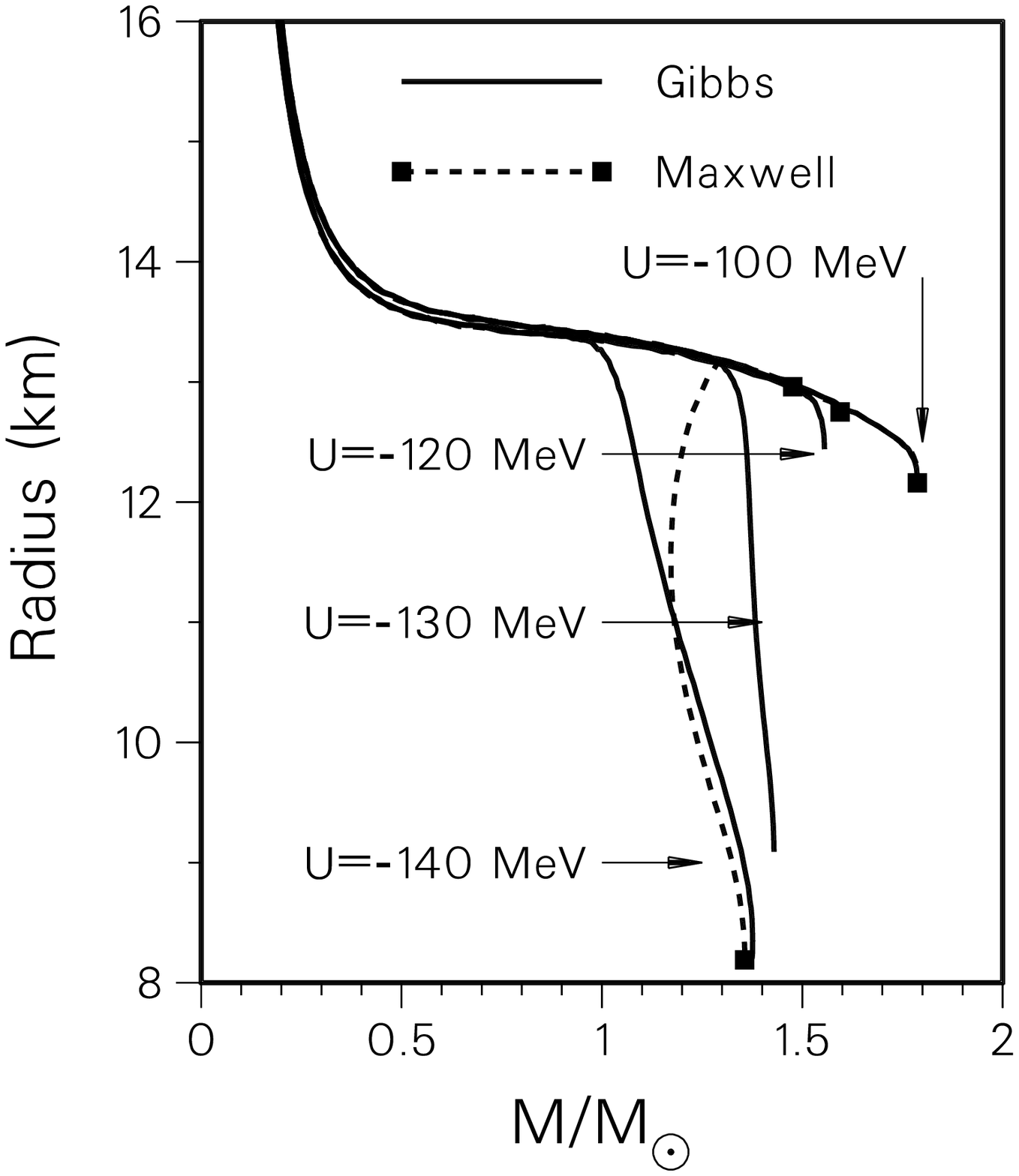}
  \caption{Mass-radius curve.Taken from ref.[22].}
  \label{fig:glerm}
 \end{minipage}%
 \hfill~%
 \begin{minipage}{0.49\textwidth}
  \epsfsize=0.99\textwidth
  \epsffile{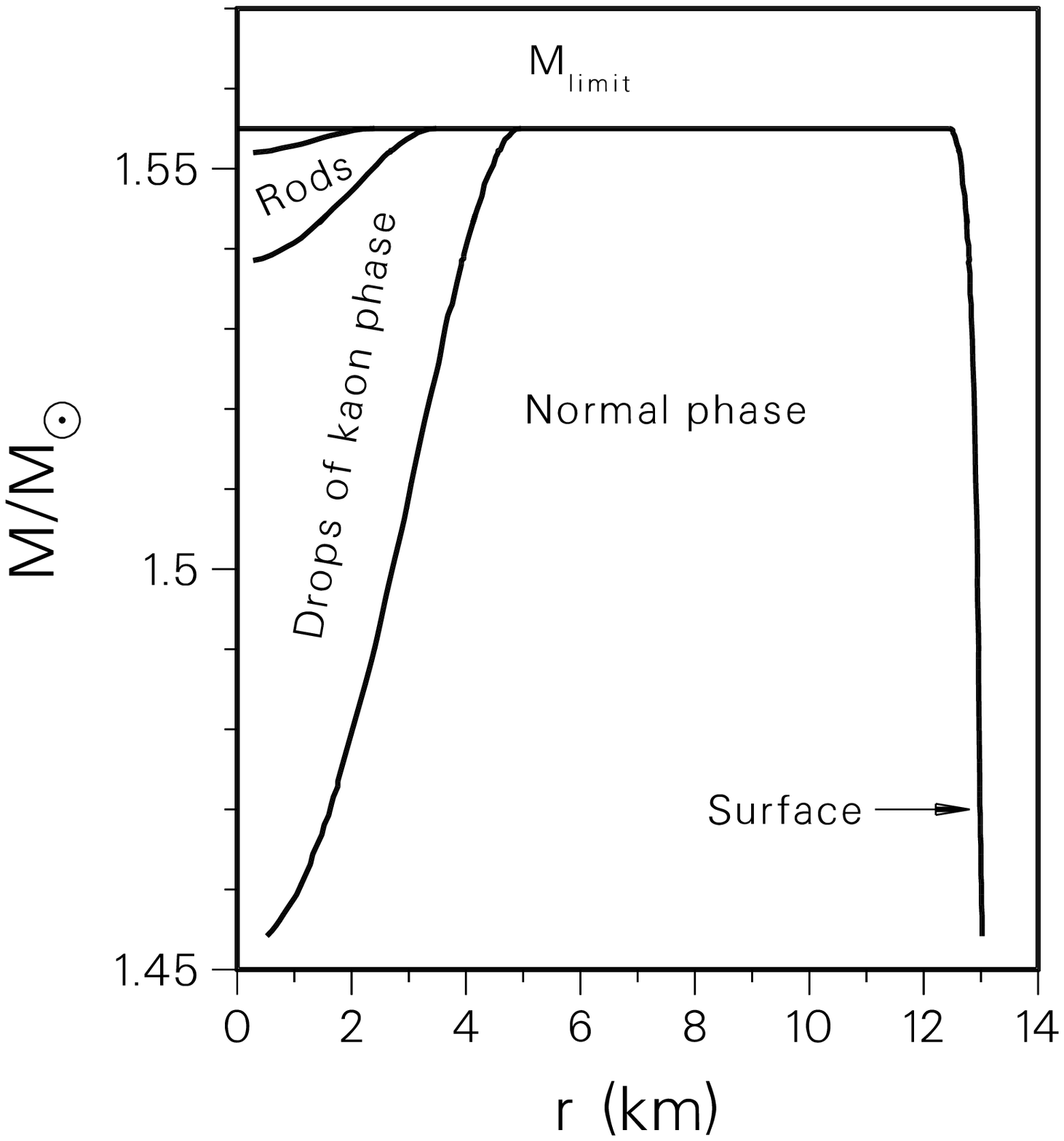}
  \caption{Interior structure of NS.
Taken from ref.[22].}
  \label{fig:glecon}
 \end{minipage}%
\end{figure}

NS structure can be studied with the EOS in Fig.\ref{fig:gleeos}.
Fig.\ref{fig:glerm} shows mass-radius curve.
We can find that
the gravitationally unstable region in the case of the
Maxwell construction
disappears in the case of the Gibbs conditions.
Here it is to be noted that
the remarkable change appears
for NS with comparably small mass (large radius)
and behavior of heavy NS is little changed.
Then our discussion about the delayed collapse
in Sect.\ref{sec:dc} will not be modified
so much.

Fig.\ref{fig:glecon} shows the interior structure
of NS in the case of the Gibbs conditions.
The mixed phase may exist largely in the cores of NS.
For a NS with $1.5M_\odot$,,
for example,
a large droplet phase exists
and its radius is about $3$[km].

\subsection{Summary for mixed phase}
In this section
we have addressed the interesting issue about the mixed phase.
The problem of the mixed phase
seems to have two important aspects.
One is the change of bulk properties of NS (mass, radius and so on.)
through the modification of EOS
and the other is of interior structure of NS
due to the appearance of large extent of the mixed phase.
The latter, especially,  may
give many implications to astrophysics.
For example,
Reddy et al.\cite{Reddy} discussed
the coherent scattering of neutrinos
in the mixed phase
and they found that
the mean free path of neutrinos
becomes smaller in the droplet phase.

In the preceding discussion,
we have ignored the finite volume effects:
surface and Coulomb energy.
They cannot be neglected
to study the realistic EOS
and they are expected to
prevent the occurrence of the mixed phase\cite{heiselQM}.
Then the realistic EOS may exist
between the EOS in the Gibbs conditions and the Maxwell construction.

\section{Magnetars and Strange Stars}\label{sec:stmag}
The magnetars are NS with strong magnetic fields
($B\sim 10^{14-15}$G derived from the $P$-$\dot{P}$ curve)
and anomalous X-ray pulsars (AXP)
and two pulsars in soft-gamma-ray repeaters (SGR)
are known as the candidates\cite{magnetar}.
The strong magnetic fields
are out of the scaling law between radius and magnetic field
$R^2B\simeq const$,
which can explain the magnetic fields
of other stars(See Table \ref{table:mag}).
\begin{table}
 \begin{center}
  \begin{tabular}{ccccc}
	& Sun & White Dwarf & NS & Magnetar\\
	$B$[gauss] & $10^{3}$ & $10^{8}$ & $10^{12}$ & $10^{15}$\\
	$R$[cm] & $10^{10}$ & $10^{8}$ & $10^{6}$ & $10^{6}$
  \end{tabular}
 \end{center}
 \caption{Magnetic fields and radii.}
 \label{table:mag}
\end{table}
Then there is a possibility
that they may originate from strong interaction.
Recently there has been proposed on idea that
the magnetars may be strange stars
with complete spin alignment\cite{TTst}.

Once $u$,$d$,$s$-quark matter is formed,
it may be more stable than normal nuclear matter
at lower densities.
Distinguished from the usual NS with core of quark matter,
compact stars consisting mainly strange quark matter
with or without thin crust
are called strange stars.

As a very simple estimation,
using non-interacting massless quarks in bag
at $T=0$,
energy per baryon of $u$,$d$-quark matter $\varepsilon_{u,d}$
and of $u$,$d$,$s$-quark matter $\varepsilon_{u,d,s}$
can be estimated.
\begin{eqnarray}
 \varepsilon_{u,d} &\simeq& 934 {\rm MeV} \frac{B^{1/4}}{145}, \nonumber \\
 \varepsilon_{u,d,s} &\simeq& 829 {\rm MeV} \frac{B^{1/4}}{145}, \nonumber
\end{eqnarray}
with bag constant $B$.
Then we can easily find that
the $u$,$d$,$s$-quark matter is more stable
than $u$,$d$-quark matter and nuclear matter.
Of course we can use more complex model,
with the one-gluon exchange and with heavy strange quark,
and the result is similar\cite{Madsen}.
A natural question may appear:
if this is right,
why nuclei consists of nucleon.
But there is no contradiction
because
there is surface effect
for light nuclei
and 
no strange quark exists for heavy nuclei.
A few strange quark in nuclear matter
are not stable
and it is almost impossible
to produce enoughly many strange quarks
at the same time
through the higher-order weak interaction.

Based on the idea of strange stars,
Tatsumi studied the possibility of
magnetization of strange quark matter
with relativistic one-gluon exchange interaction\cite{TTst}.
He concluded that
magnetized strange quark matter may be stable
in the low-density-region.
At $\rho\sim 0.1$fm$^{-3}$
magnetization may occur in strange stars
and can produce $B\sim 10^{15-17}$.
He just suggested the possibility
and realistic calculation(e.g. Hartree-Fock)
is the future problem.

\section{Concluding Remarks}
We concentrated on 3 topics in this paper.
We discussed the evolution of PNS
from the view point of the nuclear theory
(in the static limit)
in Sect.\ref{sec:dc}
and found that the delayed collapse
is possible in the deleptonization era
due to the occurrence of kaon condensation.
In Sect.\ref{sec:mp}
we reviewed the problem of the mixed phase,
which may give important effects
to NS physics:
glitch, cooling scenario and so on.
Then an interesting idea
about the identification of magnetars
as strange stars
was introduced in Sect.\ref{sec:stmag}.
Of course
there exist many candidates of phase transitions
and related astrophysical phenomena
which we could not pick up here.
And now we recommend a good and recent review
by Heiselberg and Hjorth-Jensen\cite{heiselberg}.

At present
it is not clear what kind of phase transitions
really exist in NS.
But the experimental data on the earth
and the observation of signal from NS
will limit the candidates
and, sometime, determine what happens in NS,
we hope.

\section*{ACKNOWLEDGEMENT}
It is a great pleasure to thank
the organizers of TOURS2000
for an enjoyable meeting.
The author is grateful
to T. Tatsumi, T. Muto and T. Takatsuka
for the collaborations
and helpful discussions.

\end{document}